\documentclass[sigconf]{acmart}

\AtBeginDocument{%
  \providecommand\BibTeX{{%
    \normalfont B\kern-0.5em{\scshape i\kern-0.25em b}\kern-0.8em\TeX}}}

\usepackage{subfigure}
\definecolor{codegreen}{rgb}{0,0.6,0}
\definecolor{codegray}{rgb}{0.5,0.5,0.5}
\definecolor{codepurple}{rgb}{0.58,0,0.82}
\definecolor{backcolour}{rgb}{0.95,0.95,0.92}
\definecolor{bluepigment}{rgb}{0.2, 0.2, 0.6}
\newcommand\enquote[1]{``#1''}

\copyrightyear{2022}
\acmYear{2022}
\setcopyright{acmcopyright}

\setcopyright{acmcopyright}\acmConference[Preprint]{Print}{May 2022}{Oxford, UK}
\acmBooktitle{Preprint}

\acmPrice{15.00}
\acmDOI{10.1145/1122445.1122456}
\acmISBN{978-1-4503-XXXX-X/18/06}

\usepackage{amsmath}
\usepackage{enumerate}

\usepackage{listings}

\definecolor{codegreen}{rgb}{0.36, 0.54, 0.66}
\definecolor{codegray}{rgb}{1,1,1}
\definecolor{codepurple}{rgb}{0.58,0,0.82}
\definecolor{bluepigment}{rgb}{0.2, 0.2, 0.6}
\definecolor{amethyst}{rgb}{0.6, 0.4, 0.8}
\definecolor{backcolour}{rgb}{1,1,1}
\lstdefinestyle{mystyle}{
    backgroundcolor=\color{backcolour},   
    numberstyle=\tiny\color{codegray},
    basicstyle=\ttfamily\footnotesize,
    breakatwhitespace=false,         
    breaklines=true,                 
    captionpos=b,                    
    keepspaces=true,                 
    numbers=left,                    
    numbersep=5pt,                  
    showspaces=false,                
    showstringspaces=false,
    showtabs=false,                  
    tabsize=2
}
\begin{document}

\title{Imagining, Studying and Realising A Less Harmful App Ecosystem}

\author{Konrad Kollnig}
\author{Siddhartha Datta}
\author{Nigel Shadbolt}
\email{{firstname.lastname}@cs.ox.ac.uk}
\affiliation{%
  \institution{Department of Computer Science, University of Oxford}
  \streetaddress{Parks Road}
  \city{Oxford}
  \postcode{OX1 3QD}
  \country{United Kingdom}
}
\renewcommand{\shortauthors}{Kollnig, Datta, and Shadbolt}

\begin{abstract}
\textbf{This a preprint of work-in-progress.}

Desktop browser extensions have long allowed users to improve their experience online and tackle widespread harms on websites.
So far, no equivalent solution exists for mobile apps, despite the fact that individuals now spend significantly more time on mobile than on desktop, and arguably face similarly widespread harms.

In this work, we investigate \textit{mobile app extensions}, a previously underexplored concept to study and address digital harms within mobile apps in a decentralised, community-driven way.
We analyse challenges to adoption of this approach so far, and present a ready-to-use implementation for Android as a result of significant and careful system development.
Through a range of case studies, we demonstrate that our implementation can already reduce (though not completely eliminate) a wide range of harms~--~similarly as browser extensions do on desktops.

Our method provides a versatile foundation for a range of follow-up research into digital harms in mobile apps that has not previously been possible, given that browser extensions have long been a fruitful foundation for research studies on desktops.
In other words, our system tries to address the gap of a focus on desktop interventions in previous research.
\end{abstract}



\maketitle

\section{Introduction}
\label{sec:introduction}
\begin{figure*}
    \centering
    \subfigure[Step 1: User selects an app.]{\includegraphics[width=0.28\linewidth]{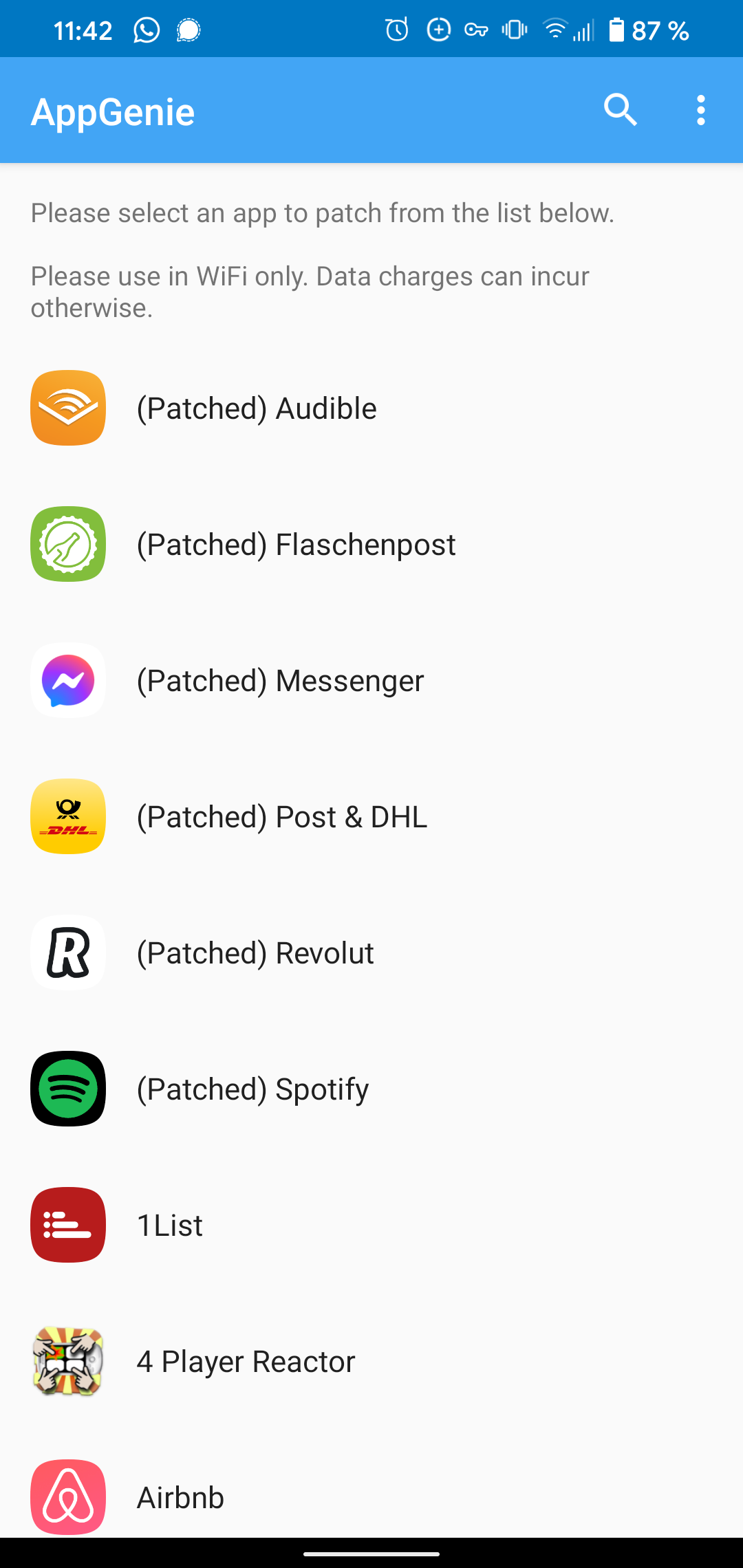}}
    \subfigure[Step 2: User selects extensions.]{\includegraphics[width=0.28\linewidth]{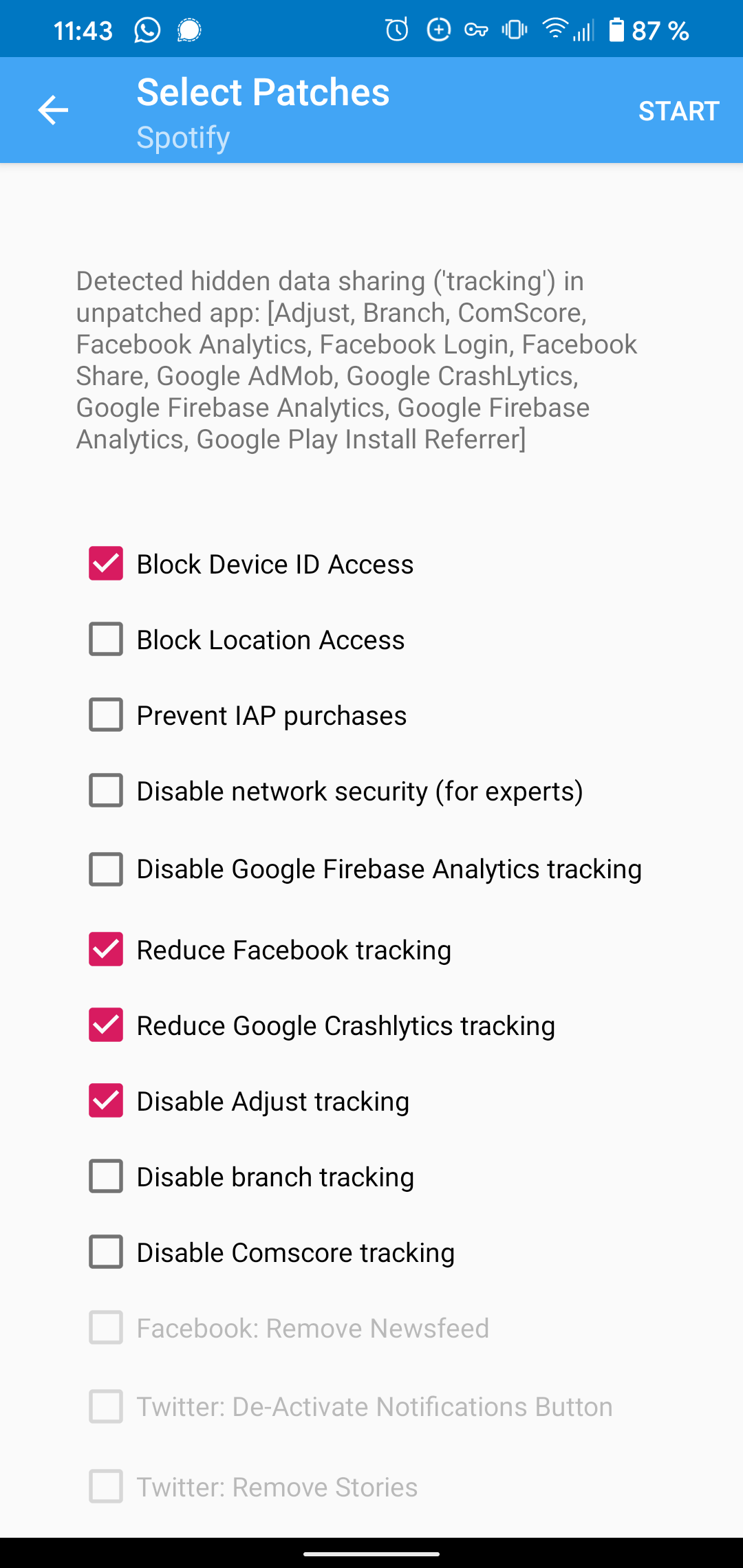}}
    \subfigure[Step 3: User installs extended app.]{\includegraphics[width=0.28\textwidth]{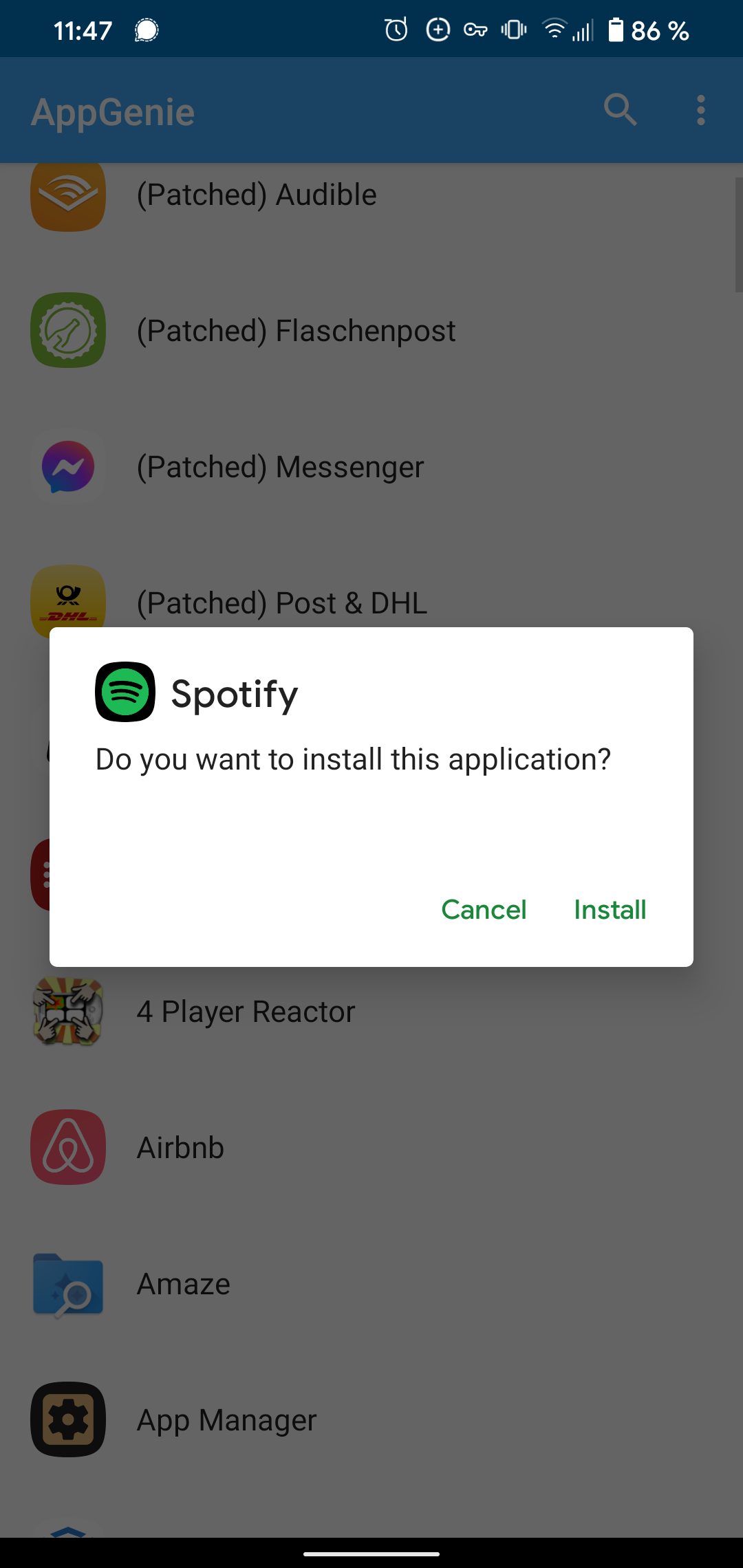}}
    \caption{Choosing and enabling app extensions in three simple steps. Our ready-to-use Android implementation enables non-expert users to reduce harms within apps. This, in turn, enables researchers to study the usefulness of such interventions easily, as well as the impact of mobile harms on individuals. For instance, the Spotify app contains an array of tracking functionality (automatically detected by our implementation), which can be removed from the app with our tool. Further extensions are available (e.g. removing \enquote{stories} bar from the Twitter app to make the app less distractive and increase user autonomy, or to make mobile apps more child-friendly), but not all of them apply to the Spotify app (greyed out); more in Section~\ref{sec:casestudies}}
    \label{fig:steps}
\end{figure*}

The WWW was invented in 1989 on an open, equal and universal premise to share and access information across computers.
More than thirty years later, these original values are challenged by a digital ecosystem that is dominated by a few gatekeeper companies.
The predominant business model in the digital ecosystem is one that relies on the mass-scale collection of personal data, to create fine-grained profiles of individuals and serve personalised ads.
User choice is often limited in 1) how their data is used, and 2) in how the underlying software architecture is designed.

The harms arising from the current design of the app ecosystem are multi-faceted, growing and widespread.
They can include the loss of privacy~\cite{binns_third_2018,zimmeck_maps_2019}, dark patterns~\cite{nouwens_dark_2020,10.1145/3173574.3174108,10.1145/3334480.3381070}, digital distraction~\cite{lyngs_self-control_2019,kovacs_thesis,lyngs_i_2020,lukoff_youtube}, digital inclusion~\cite{wu_design_2019,rieger_sinders_2020}, discrimination and algorithmic fairness~\cite{benjamin_race_2019,binns_how_2021} and harmful social media content~\cite{hmgov}.
Addressing these harms is particularly relevant in a world that is recovering from the COVID-19 pandemic, and is spending more time on their mobile devices than ever before~\cite{appannie_mobile_stats_2022}.

Motivated by the pervasiveness and ubiquity of online harms, a broad range of countermeasures is being proposed and developed.
For years, the industry has advocated self-regulation as an appropriate remedy. The recent release of the Facebook Papers suggests that this approach has not prevented immense harm caused.
Existing legal instruments, such as data protection or competition laws, are deemed to be inflexible and take too much time to keep up with the quickly changing digital world~\cite{binns_dissolving_2020,lynskey_grappling_2019,wu_curse_2018}.
This is why, increasingly, new regulation is being developed, such as the Online Harms Bill in the UK, the Digital Services Act in the EU, and most recently the Open App Markets Act in the US.
Whether or not such new legal remedies will overcome the shortcomings of past regulation and meaningfully improve the technical nature of digital systems remains to be seen.

A key challenge remains the reliance of individuals on the goodwill, interest, and ability of app developers, platform gatekeepers, regulators and civil society to do so.
What if, instead, end-users were given the opportunity to choose for themselves what is best for them, rather than solely rely on the discretion of others, particularly gatekeepers that are driven by sometimes very different values and interests than consumers?
This is why this work will explore technical interventions, which could enhance and complement other existing and emerging forms of regulation.

This work makes two main contributions:
\begin{enumerate}
	\item \textbf{Conceptual Contribution:} We introduce the community-driven \textit{concept} of mobile app extensions; discuss the reasons why a similar system exists in desktop browsers, but not mobile devices; and explore how such a system might be helpful for research studies on harms in mobile apps as well as end-user regulation of digital technologies.
	\item \textbf{Technical Contribution:} We design and implement a versatile and ready-to-use \textit{implementation} of mobile app extensions, including an Android app. This system is the result of significant and careful system design and development. As we will discuss, our implementation already allows end-users to reduce various harms inside the apps they regularly use, by making targeted modifications to such apps~--~similar as browser extensions do with websites.
\end{enumerate}
We evaluate our work through a range of case studies, and share the code of implementation.

Mobile app extensions will not be a magical fix to \textit{all} harms in apps, and have the known limitations of browser extensions.
However, this concept might still be helpful for \textit{researchers} to study harms in mobile devices, and for \textit{end-users} to overcome such harms in their day-to-day lives.
At the moment, research studies often choose studying interventions against harms on desktop rather mobile, because no ready-to-use extension system has so far existed~\cite{lyngs_i_2020,lukoff_youtube}.
Likewise, end-users have limited means to avoid harms in those apps that they rely on in their daily lives, such as social media, shopping, and dating apps.
Our work tries to address this gap, by both investigating the concept theoretically and proposing a functional implementation for Android.

The rest of this paper is structured as follows.
In Section~\ref{sec:background}, we introduce the necessary background, in particular why extensions exist in desktop browsers, but not on mobile, and the implications of this.
In Section~\ref{sec:framework}, we introduce the concept of mobile app extensions. As part of, we derive key requirements that an app extension system should have in Section~\ref{sec:reqs} and put forward a high-level design based on these requirements in Section~\ref{sec:design}.
In Section~\ref{sec:implementation}, we introduce a sample implementation of this system design that works for Android.
In Section~\ref{sec:casestudies}, we evaluate our approach against the requirements derived in Section~\ref{sec:reqs} through a set of case studies.
In Section~\ref{sec:discussion}, we discuss the learnings from our exploration of mobile app extensions, point out the varied challenges that still lie ahead if mobile app extensions should be deemed useful by interested researchers and other individuals, and also the promises of the approach for the future.
In Section~\ref{sec:conclusions}, we summarise our work and draw conclusions. In particular, we point out that mobile app extensions will unlikely be possible without the explicit support of device manufacturers, notably Apple and Google.
We also give directions for future work.

\section{Background}
\label{sec:background}
In this section, we first motivate our work by a review of the technology of browser extensions.
We then discuss similar existing such approaches on mobile devices.

\subsection{Browser Extensions and the Importance of Autonomy}

On desktops, end-users have long been able to choose from a wide variety of browser extensions to address some of the widespread harms in the digital ecosystems.
The ability to install extensions has contributed to the early success of Mozilla Firefox over the Internet Explorer in the 2000, and has led to a vast ecosystem of extensions that improve users' browser experience.
These extensions help end-users remove malware from websites (e.g. through a tracker blocker~\cite{tcontrol,adblockers_court_2022}),
reduce unwanted distractions and dark patterns (e.g. by removing the Facebook feed and other distracting elements~\cite{lyngs_i_2020,xda}), make it easier for disadvantaged users to participate by allowing them to make the web more accessible (e.g. through browser extensions that render any displayed text more readable for dyslexics~\cite{beeline,helperbird}), and
help platform workers to negotiate better work conditions~\cite{irani_turkopticon_2013}.
Nowadays, web extensions are nothing more than HTML5 applications that are distributed to end-users through extensions stores, notably the Google Chrome Web Store and the Mozilla Firefox Add-ons Portal.
With the WWW being built around open technologies, making browser extensions follow this example seems like a natural design choice to ensure the success of the system.
The distribution through extensions stores additionally provides a protection layer against misuse of the browser extension functionality.

The story of browser extensions is not without drawbacks.
With an ever-growing market share in browser engines, Google has been making key changes to its Chromium project that will limit the current abilities of browser extensions in the future.
One of the most recent examples is the ongoing introduction of Manifest V3 that is making the development of ad and tracking blockers much more difficult because Google artificially limits the number of blocking rules that such blockers can specific~\cite{borgolte_understanding_2020}.
This is why competition authorities around the globe, particularly in the Competition and Markets Authority in the UK, are exploring what negative impacts that Google's control over browser, smartphone, and advertising technology might have on consumers~\cite{competition_and_markets_authority_online_2020,competition_and_markets_authority_online_privacy_sandbox}.

Bongard-Blanchy et al.~\cite{10.1145/3461778.3462086} and Gray et al.~\cite{10.1145/3479516} verified through user studies that users are often not aware of the presence of dark patterns, and experience \enquote{feeling manipulated.}
There is often a steep power imbalance between platforms 
and end-users. In analysing user autonomy, Susser et al.~\cite{suss19} raised concerns on how digital surveillance and tracking of user data can be used to discover and manipulate end-user vulnerabilities, or how personalisation of platforms exploits our individual decision-making and deprives users of a shared experience.

There are a range of limitations to current implementations of browser extensions.
First, their scope of applicability is limited and not all harms within app can be mitigated.
Many web applications are hosted in the cloud, but browser extensions run on users' devices and not the cloud-based backend systems.
Second, there have been reported cases of malicious browser extensions in the past.
While the operators of extension stores were usually quick to address these threats, they can still have a sizeable impact on the lives of some individuals.
Such threats are, however, inherent to any form of app stores, and are unlikely to be overcome in this model of software distribution.
Third, there is an ongoing debate around what kinds of browser extensions are permissible.
This especially relates to browser extensions that tinker with the revenue flows of websites, i.e. ad blockers.
The permissiveness of ad blockers has been being challenged for about a decade by online publishers (and Axel Springer in particular) in German courts. The courts, however, have consistently reasserted users' principle right to make such modifications to their online experience~\cite{adblockers_court_2018,adblockers_court_2022}.
When it comes to non-revenue-related extensions, browser extensions are much less controversial.
This is why we will focus on extensions other than ad blocking in this paper.

The limitations of browser extensions aside, they can still make a tangible improvement to the browser experience of individuals.
Moreover, they have also, in the past, facilitated a broad range of research studies~\cite{lyngs_i_2020,kovacs_thesis}.
This is why an equivalent of browser extensions on mobile might provide similar benefits in apps, which are some of the most widely used pieces of technology.
We will thus explore the concept of \textit{mobile app extensions} in the rest of this paper~--~a concept that has not previously been thoroughly discussed.

\subsection{Mobile App Extensions: The Current Landscape}
\label{sec:appextensions}

Mobile ecosystems emerged as locked-up ecosystems that isolate different apps for security and usability purposes and are largely incompatible with open web technologies.
While on the web, users used to be able to install browser extensions, similar technology has not been allowed on mobile devices.
There is no technical reason why this should not be the case.
Apple has already introduced such with iOS 15, but limits them to its Safari browser.
Google has long been criticised for not extending the extension functionality from its desktop browser to its mobile browser (which would be technically very simple, since the Google Chrome mobile and desktop browsers have the same code base, and mobile Safari~--~using a similar code base~--~has already implemented this).

While having had mixed success so far, there exist a range of solutions that try to extend the functionality of mobile apps.
Methodologically, these solutions either
1) modify the operating system (e.g. Cydia Substrate~\cite{cydia}, Xposed Framework~\cite{xp}, ProtectMyPrivacy~\cite{agarwal_protectmyprivacy_2013}, or TaintDroid~\cite{enck_taintdroid_2010}),
2) modify apps directly (e.g. 
GreaseDroid~\cite{10.1145/3411763.3451632},
Lucky Patcher~\cite{lp}, apk-mitm~\cite{apkm}, Joen et al.~\cite{jeon_dr_2012}, DroidForce~\cite{rasthofer_droidforce_2014}, RetroSkeleton~\cite{davis_retroskeleton_2013}, SRT AppGuard~\cite{garcia-alfaro_appguard_2014}, I-ARM-Droid~\cite{Davis12i-arm-droid:a}, Aurasium~\cite{Aurasium}, Objection~\cite{objection}), or
3) use System APIs (e.g. VPN-based ad blockers, such as AdGuard~\cite{ag} and TrackerControl~\cite{tcontrol}, or UI overlays, such as DetoxDroid~\cite{detoxdroid}, Gray-Switch~\cite{grayswitch} the Google Accessibility Suite~\cite{accessibility}, and GreaseTerminator~\cite{greaseterminator}).

All of these solutions come with limitations.
While modifying the operating system can in principle make arbitrary modifications to the behaviour of apps, these usually rely on device vulnerabilities and are a moving target for device manufacturers. Operating system modifications can pose security risks, potentially void warranty, and are usually infeasible for non-expert users.
By contrast, the use of System APIs might often be the most straightforward approach for a non-expert user, operating in the familiar space of the user's own smartphone. This also poses a major limitation because only what is permitted by the smartphone operating system can be realised.
For instance, the removal of the newsfeed from the Facebook app has not been accomplished through System APIs, though this can be an effective way to get more control over one's Facebook use~\cite{lyngs_i_2020}.

In app modification on Android, some transformation is applied to an app used by the user, e.g. the removal of the newsfeed from the Facebook app.
App modification offers ease of use (installing custom apps on Android is supported by the operating system), and allows for~--~theoretically~--~arbitrary modifications of a provided app (only limited by the constraints of the operating system). As such, it combines benefits of system modification and System APIs.
Despite this, there exist hardly any solutions used in practice using app modification.
Exceptions are the app cracking tool Lucky Patcher~\cite{lp}, the privacy tool SRT AppGuard~\cite{garcia-alfaro_appguard_2014}, and \textit{apk-mitm}~\cite{apkm} and \textit{objection}~\cite{objection}, both for app security research.
GreaseDroid~\cite{10.1145/3411763.3451632} is another exception, but only aimed at dark patterns and was an early research prototype not ready for use in practice.
Some developers have published modified versions of popular apps, including Facebook~\cite{xda} and YouTube~\cite{vanced} (removing ads and other distracting functions). Unfortunately, such modified apps rely on the continued support of the developers, may break over time (e.g. in case of server-side updates), and exist only for a few select apps.

Many of the current app extension tools are neither allowed on the official app stores nor supported by the gatekeepers.
As such, these tools \textit{currently} do not undergo the usual review process and can pose unexpected risks to their users.
Unfortunately, these technologies have been targeted by the gatekeepers of mobile ecosystems
Most recently, the main developer behind Magisk was hired by Google and the developer had to remove core functionality of Magisk as a result of obligations imposed by this new employer.
Moreover, Google currently bans any apps from its Play Store \enquote{that interfere with ... other apps}~\cite{playstore_interference}.
This has in the past been the foundation of Google’s ban of the popular privacy and anti-malware software Disconnect.Me from the Play Store in 2014.

\begin{figure*}
	\centering
	\includegraphics[width=\textwidth]{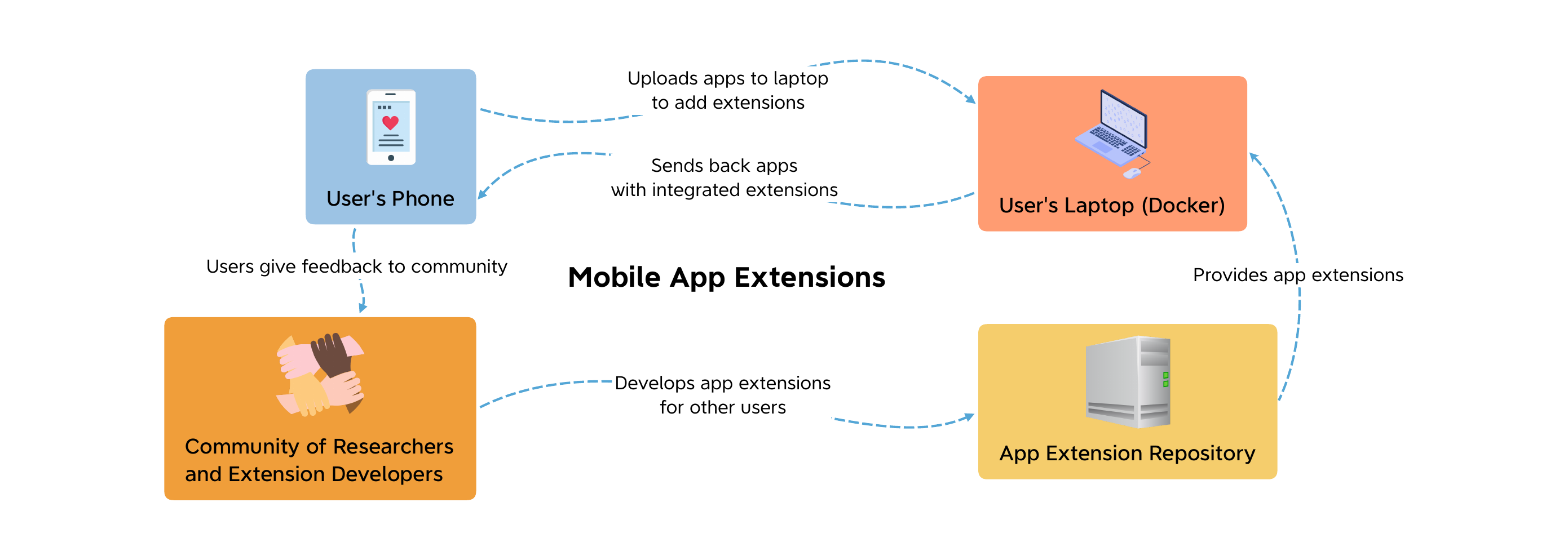}
	\caption{Overview of our current \textit{implementation} of Mobile App Extensions. Compared to browser extensions, our implementation currently relies on an external laptop to apply extensions to mobile apps. In the future, other approaches, particularly the explicit support by device manufacturers, might overcome these current limitations.}
	\label{fig:overview}
\end{figure*}

\section{Framework: Addressing Harms through Mobile App Extensions}
\label{sec:framework}
The previous Section~\ref{sec:background} reviewed past work and highlighted that a system similar to browser extensions for mobile might be helpful for end-users and researchers alike, in studying and mitigating mobile harms.
Based on this literature review, we first derive key system requirements that a mobile app extension system should fulfil.
We then derive a high-level system design based on these requirements.
Lastly, we introduce an implementation of this system design for Android that is meant to fulfil the key requirements.

We will evaluate in the subsequent Section~\ref{sec:casestudies} whether and to what extent our Android \textit{implementation} of mobile extensions actually fulfils the necessary system requirements through a range of case studies, and discuss the learnings for future work and the \textit{concept} of app extensions in the Discussion Section~\ref{sec:discussion}.

\subsection{System Requirements}
\label{sec:reqs}

From the literature review in the previous section, three requirements emerge that an app extension system should have.
A mobile app extension system should fulfil these requirements in order to allow users without expert knowledge to address harms within mobile apps:

\textbf{Requirement 1: Versatility.}
As noted in the previous section, digital harms are incredibly wide-range and individual.
Therefore, a key requirement of our system should be the ability to cover a wide range of harms (and ideally across a wide range of apps).

\textbf{Requirement 2: Usability.}
Usability encompasses both those that use mobile extensions and those that develop them.
Users of extensions should be able to install any extension easily on their digital device.
Developers of extensions should be able to develop extensions against digital harms fairly easily.
The simplicity of development has greatly contributed to the success of browser extensions, which rely on JavaScript and other web technologies for development.
However, at this stage of development, similar simplicity might be unrealistic, given that the technology behind browser extensions has been developed over more than a decade and has been evaluated with billions of users.

\textbf{Requirement 3: Security.}
As discussed above, the distribution of software from third-party vendors can raise security problems. This is why it is important to address this issue in any extension system.
Security is usually addressed through a thorough review process.

\subsection{System design} 
\label{sec:design}

App modification (as discussed in Section~\ref{sec:appextensions}) bears potential as a useful method to address and study existing and emerging harms in mobile ecosystems.
At the moment however, this method remains inaccessible for many end-users and researchers that lack the necessary programming knowledge, and constitutes a potentially missed opportunity to study and reduce harms.
To address this, we propose a high-level system design on a decentralised, community-based approach. This system aims to fulfil the key system requirements from the previous section that a mobile app extension system should fulfil, according to the literature review.
In the following subsection, we then discuss how we concretely implemented this on Android.
We visualise our implementation in Figure~\ref{fig:overview}.

\textbf{Stakeholders.} We envisage an extension ecosystem for mobile apps, consisting of \textit{extension developers} (e.g. researchers or other developers) and \textit{end-users} (e.g. study participants and other users without expert knowledge).
Extension developers are individuals with programming skills that develop \textit{extensions}.
An extension, in turn, contains instructions on how to modify certain apps.
End-users can choose from a \textit{repository of extensions} to modify their apps as they want.
The more users (both extension developers and end-users) use the system, the higher the likelihood that even end-users without expert skills will find the extensions they need.
Structurally, our system is similar to browser extensions, a concept that has proven fruitful in desktop browsers and the creation of many features that are now embedded by default into these browsers (such as developer tools and privacy technologies).
By focusing on mobile, we aim to expand this idea from websites to apps.

While, ideally, we would like to enable everyone to create extensions, even individuals without programming skills, many years of work around browser extensions are only slowing making this possible, so we believe it will be infeasible for a mobile extension system in the short term.
However, the development has become easier over the years, by shifting from ActiveX-based Internet Explorer and XUL-based Firefox extensions (both rather exotic software development frameworks) to WebExtensions (based on web technologies), so the development of mobile extensions might take a similar turn in future work.
At the root of the success of web extensions lie the unique features of the networked world, in which people are often willing to collaborate, and often do so for varied, non-monetary incentives, including social recognition, as seen on platforms like Wikipedia or OpenStreetMaps~\cite{benkler_wealth_2006,socialmachines}.

\textbf{System architecture.} Since app modification can be a compute-intensive task, we opt for a client/server model.
An end-user would install a dedicated \textit{extension app} on their phone, and install the server application on their laptop.
Inside this app, there are three steps: 1) selection of an app to extend, 2) selection of extensions, 3) application of the extensions and re-installation.
In the last step, the extension app uploads the selected app to the server/laptop, and the server applies the selected app extensions to the uploaded app (i.e. this implements an app modification method, as discussed in Section~\ref{sec:appextensions}).
When the server has finished applying the selected extensions, the server sends back the resulting extended app to the user's phone, and the user is prompted by extension app to install the extended app (Step 3 in Figure~\ref{fig:steps}).
Figure~\ref{fig:steps} shows what this looks like in our implementation (more details on this will follow in the next subsection).

The focus on extension scripts, which enable the sharing of extensions with other individuals, decouples the development of extensions from the process of applying extensions to an app (thereby avoiding the previous sharing of extended apps on online forums, with uncertain security, see Section~\ref{sec:appextensions}), and contributes to the usability of our system by end-users (i.e. Requirement 2 in Section~\ref{sec:reqs}).
All an end-user needs in order to extend an app and reduce harms within it is the right extension script. The application of extensions happens with a few clicks, as seen in Figure~\ref{fig:steps}.
Extensions promise high versatility of our system (i.e. Requirement 1 in Section~\ref{sec:reqs}), provided that these extensions themselves allow for a broad range of modifications of apps.

\textbf{Review Process.}
To ensure the security of our system (i.e. Requirement 3 in Section~\ref{sec:reqs}), we envisage a review system similar to the one currently used in current app and browser extension stores.
It is beyond the scope of our work to develop such a system from scratch.
Instead, we envisage that mobile extensions might eventually be distributed through the existing software distribution process in mobile ecosystems, as we discuss in our Conclusions in Section~\ref{sec:conclusions}.

\subsection{System implementation}
\label{sec:implementation}

We now provide details on how we implemented the extension ecosystem described in the previous sections.
An overview of our implementation is shown in Figure~\ref{fig:overview}.
We will also make our implementation publicly available, since it represents the result of significant and careful system design and engineering.

\subsubsection{Application of Extensions}

Our implementation of the extension app from the previous section is a simple Android app that extracts another app selected by the end-user from the smartphone storage in the *.apk format, and uploads this *.apk file to the server on the users' laptop.
While our system could directly fetch the latest version of the app from Google Play Store, we opt for direct extraction from the user's phone to avoid potential violations of the Terms \& Conditions of the Google Play Store.
Once all extensions are applied, our app guides the user through installing the extended app on their phone.
The extended app is signed with a user-generated certificate to ensure the authenticity of the app on the Android system.
Since the extended app has a different signature than the original app, the re-installing involves uninstalling the current app on the user's phone, due to Android's security model.
This breaks compatibility with updates from the Google Play Store, but, due to the simplicity of the extension process, extensions can be easily applied to any app updates.
By installing the existing Aurora app store (i.e. an open-source client for the Google Play Store), a user could easily retrieve notifications about the availability of app updates.

In our implementation,
we use a Ubuntu-based \texttt{node.js} server running in a Docker environment to facilitate the app extension.
This Docker environment allows end-users to install the server on their laptop with only a few clicks, and on any operating system since Docker exists for Windows, Linux and macOS.
Future implementations of our system may reduce the load on end-users even further.
For now, the use of Docker avoids compatibility issues between different operating systems, and makes the extension process for end-users rather simple and streamlined.
Inside the server, we use the popular \texttt{apktool} framework to decode (and eventually re-code) every uploaded app.
The results of this decoding are Android layout files (*.xml), app bytecode (*.smali), and other app resources that can then be modified.

After applying all extensions to the decoded apps, we use \texttt{apktool} to create a working *.apk file again (\textit{re-coding}), sign the result with Google's \texttt{apksigner}, and compute the differences between the new and old app with \texttt{ApkDiffPatch}.
Eventually, the server sends back only these computed differences to the end-user's device, instead of having to share an entire *.apk file.
The whole process is done within a few minutes; applying the extensions happens in seconds; decoding and re-coding the *.apk files are the most compute-intensive parts, and we hope to address this in future work.

We keep a copy of the original, unmodified app so that users can revert back, or create a new version of the app with a different set of extensions.
Future work should address how to disable extension on-the-fly instantaneously without the need to re-run the whole process.
For now, user data is kept when installing a new version of the app with different extensions because the same certificate would be used for signing the app.

\subsubsection{Development of Extensions}

\begin{figure*}
	\includegraphics[width=\textwidth]{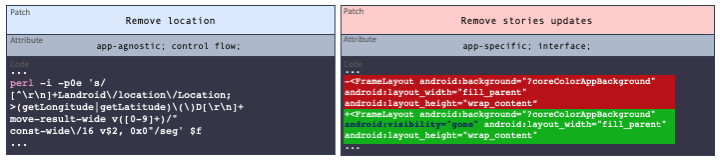}
\caption{Left: An example of an \textit{app-agnostic} extension -- a Perl script to prevent apps from accessing the longitude and latitude of the current physical user location. Right: An \textit{app-specific} extension -- a diff script to remove the \textit{Stories} bar from the Twitter app.}
\Description{Extensions describe how to modify decoded Android apps. The figures show two examples of such modifications, that is, how the app code needs to be changed in order to incorporate two example extensions.}
\label{fig:code}
\end{figure*}

In our implementation, extensions are no more than \texttt{bash} scripts. These scripts contain instructions on how to modify the decoded Android files.
The advantage of this approach is that \texttt{bash} scripts can make use of arbitrary Linux commands to modify Android apps, and~--~theoretically~--~can change all aspects about Android apps.
As we will explore in our case studies in Section~\ref{sec:casestudies}, there are several challenges in developing of extensions against arbitrary harms in practice.
Extension developers, in turn, only need to be familiar with \texttt{bash} and Android development to modify the resources files of apps, including layouts. Additionally, to modify the code of apps, extension developers need knowledge of Android bytecode, about which there exist rich resources online.
Two examples are shown in Figure~\ref{fig:code}.

Extension developers can create extension scripts by first decoding an app with {\small \texttt{apktool}}. They then sift through the app bytecode and other resources to identify potential code modifications to reduce digital harms. If the extensions are generalisable over a number of apps, these are considered \textit{app-agnostic}, else \textit{app-specific}.
Extension developers should take measures to make their extensions robust, including compatibility with extensions and with updates to the app or operating system.
Since extensions are \texttt{bash} scripts these can be understood with few additional skills. This enables the verification of extensions by other users of the system, and helps identify potentially malicious extensions, reducing security risks through transparency by design.
We provide detailed case studies of extension development in Section~\ref{sec:casestudies}.

\section{Evaluation}
\label{sec:casestudies}
We now turn to a few case studies to evaluate our proposed implementation of mobile app extensions from Section~\ref{sec:implementation} against the requirements derived in Section~\ref{sec:reqs}: versatility, usability and security.
By evaluating our concrete \textit{implementation} of mobile app extensions, we also aim to learn general lessons for the \textit{concept} of mobile app extensions and the development of such systems in the future.

In our evaluation, we mainly focus on: 1) the applicability of mobile app extensions to a wide array of harms (Requirement 1, Versatility) and 2) study the usability for \textit{researchers} (Requirement 2, Usability), particularly the ease of extension development.
We do not assess Requirement 3, Security, because the proposed review model of browser extension and app stores is a widely accepted status quo for such a system, and it is beyond this work to establish such a store.
We neither assess Requirement 2, Usability, for \textit{end-users} because there are only three steps to the application of extensions in our implementation (see Figure~\ref{fig:steps}).
Also, the principal usability of extension systems has already been widely demonstrated on the web and in research studies leveraging browser extensions.
We plan to conduct user studies for specific use cases of app extensions, for example to increase satisfaction in Facebook or YouTube use, in future research, similar to previous research on the web~\cite{lyngs_i_2020,lukoff_youtube}. 

First, we now focus on how our implementation can help reduce dark patterns in apps and support research in digital self-control.
Second, we discuss how our implementation can help in the security and privacy domain.
Third, we briefly turn to risks in apps for children, i.e. some of the most vulnerable users of mobile apps.
We will discuss the implications from our case studies in the subsequent Discussion Section~\ref{sec:discussion}, particularly the challenges for the future if one wants to establish a functioning mobile app extension system.

\begin{figure}
	\centering
	\subfigure[Removal of Facebook news feed to reduce unnecessary distractions.]{\includegraphics[width=0.49\columnwidth]{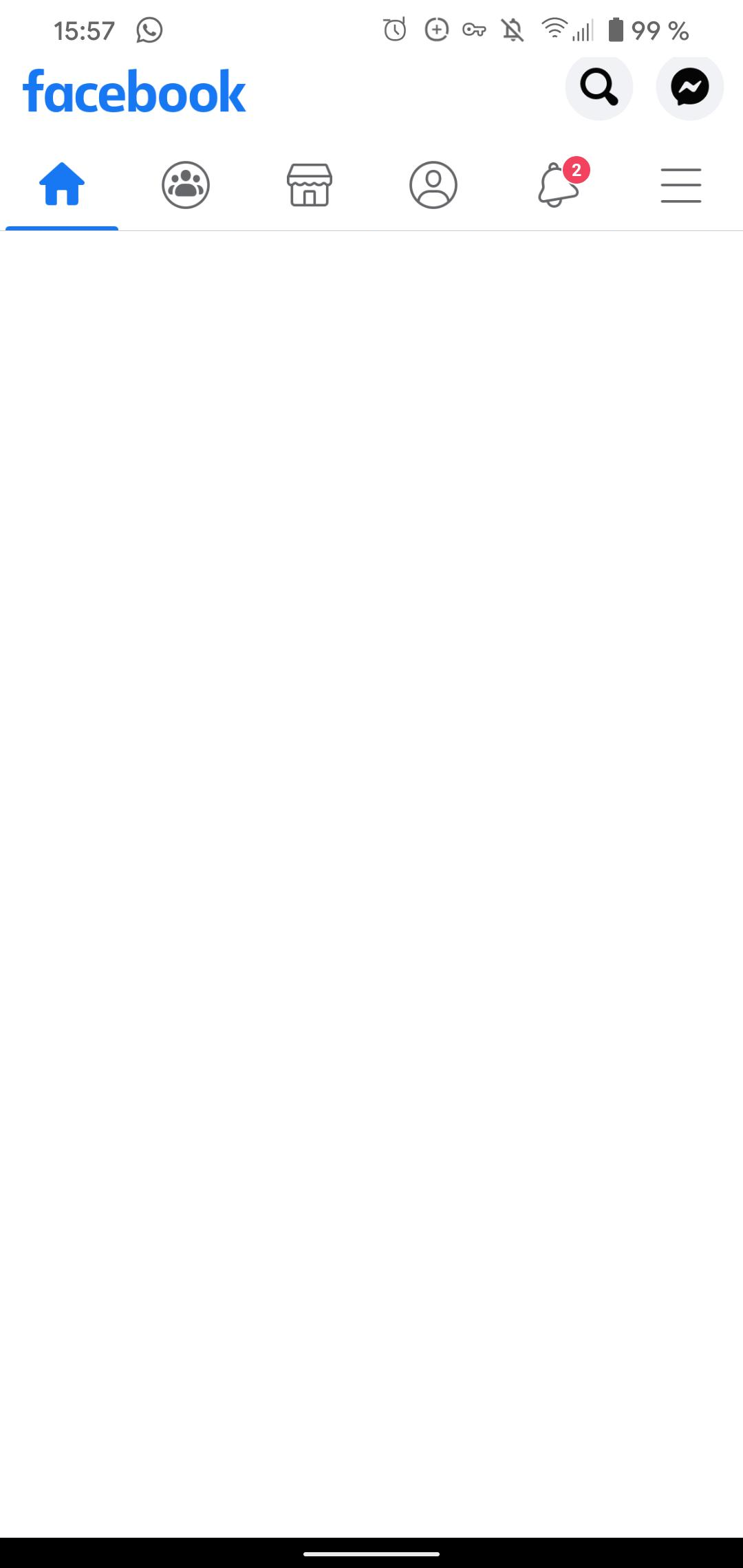}}
	\subfigure[Control condition for research studies of mobile Facebook use: added black bars.]{\includegraphics[width=0.49\columnwidth]{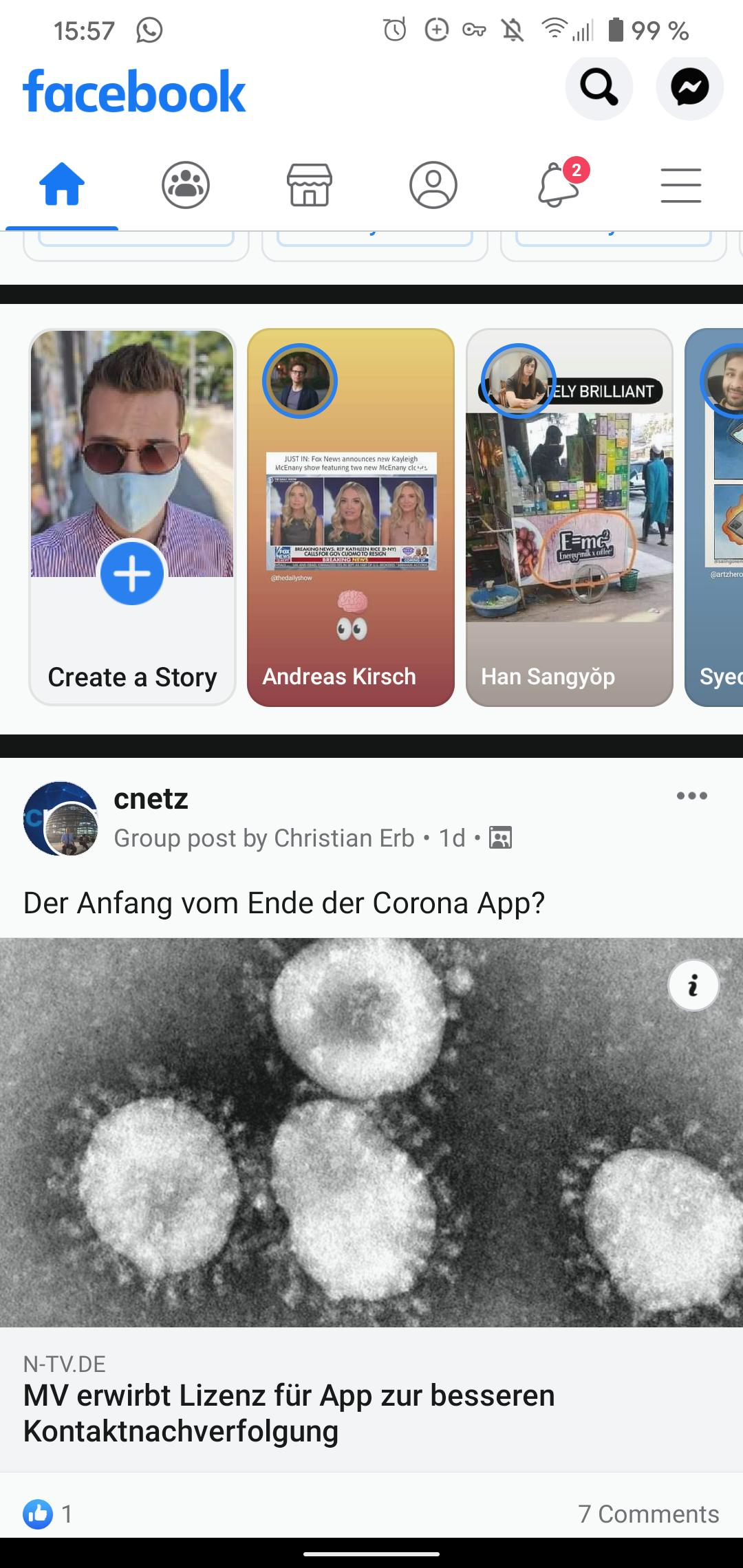}}
	\caption{
		Mobile app extensions help reduce and study a range of digital harms. One of the most commonly studied such harms in literature is the Facebook news feed that is highly engaging and can make users spend more time on the app that initially intended~\cite{lyngs_i_2020}. Past research on digital harms often tended to focus on computers, despite the fact that most individuals nowadays spend more time on smartphones. The image on the right shows a possible control condition for the study of Facebook use: added black bars.
	}
	\label{fig:facebook_newsfeed}
\end{figure}

\subsection{Digital distraction and dark patterns}

In the past years, researchers working on digital self-control have pointed out that many users find their digital devices distracting and feel like they spend too much time on them~\cite{lyngs_self-control_2019,kovacs_thesis,lyngs_i_2020,lukoff_youtube}.
Some of the research in this field has empirically studied interventions to increase self-control, and what strategies app developers use to make their apps more addictive.
However, most research focuses on desktop devices and few on mobile, due to the inherent limitations of current operating systems~\cite{lyngs_i_2020,lukoff_youtube}.
At the same time, users nowadays spend significantly more time with their mobile devices; indeed, many households (esp. of younger individuals) do not possess a computer at all.

\begin{figure}
	\centering
	\subfigure[Original Twitter app.]{\includegraphics[width=0.49\columnwidth]{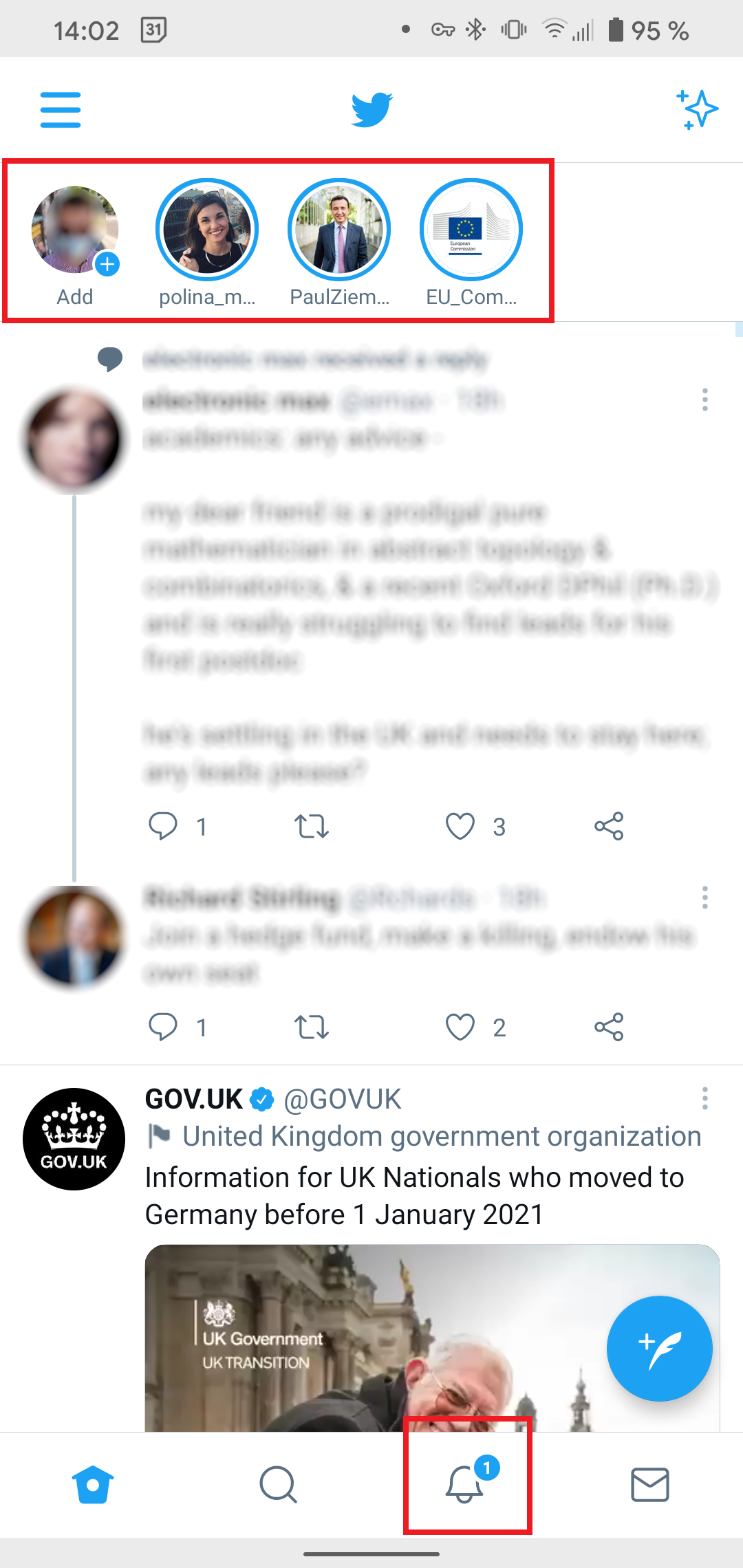}}
	\subfigure[Extended Twitter application with reduced dark patterns.]{\includegraphics[width=0.49\columnwidth]{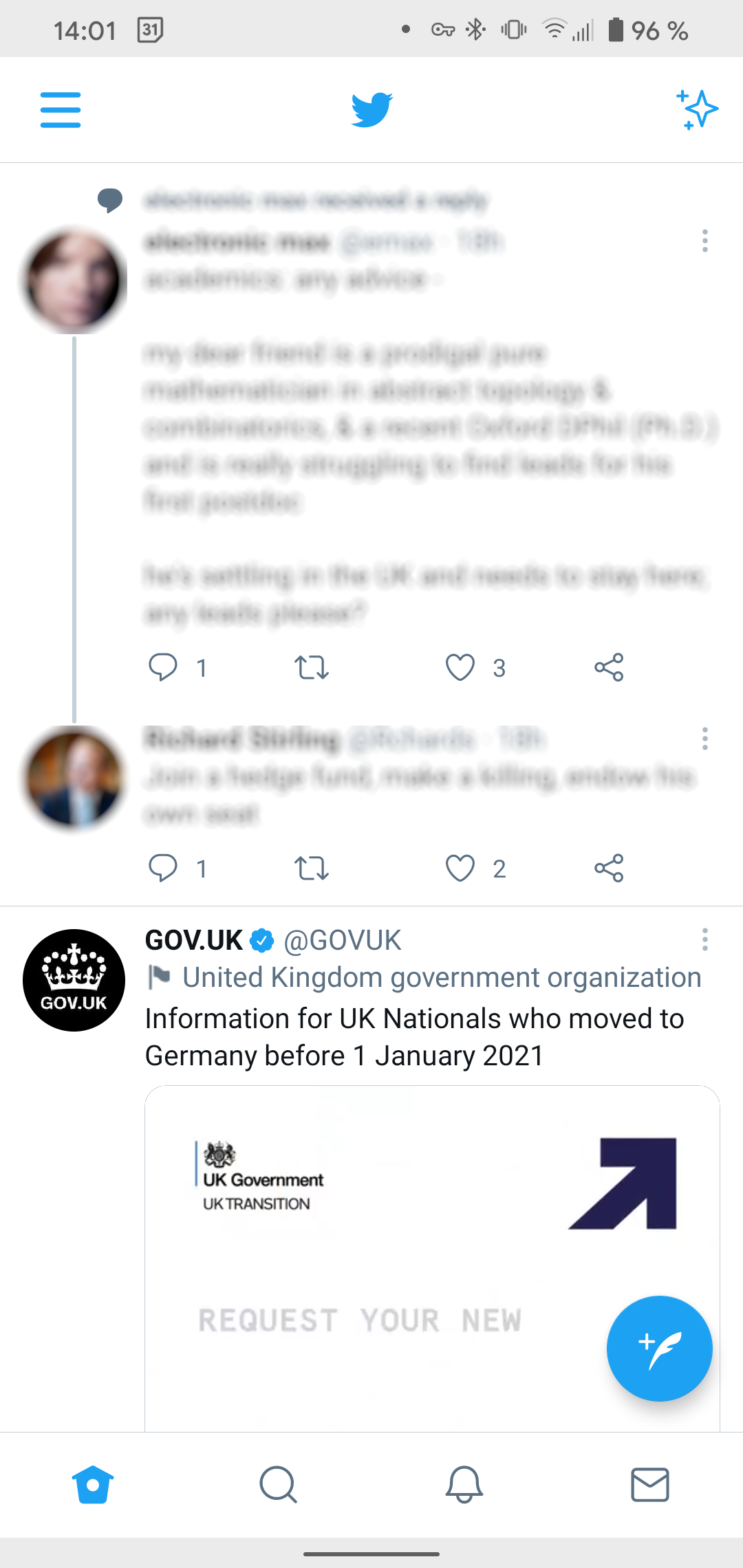}}
	\caption{
		Our method enables the removal of dark patterns (in red) in Android apps.
		Compared to the default Twitter app (left), stories and notifications have been disabled to reduce distractions in the extended version of the app (right).
	}
	\Description{The figure shows a screenshot of the default Twitter app and a screenshot of the Twitter app with two dark patterns removed (namely notification highlights and Twitter stories).}
	\label{fig:ptch}
\end{figure}

Lyngs et al.~\cite{lyngs_i_2020} found that the removal of the Facebook feed \textit{on desktop} increases users' feeling of control over their Facebook use, while reducing their time spent on Facebook.
As for the analysis of distraction in mobile apps, Lukoff et al.~\cite{lukoff_youtube} created a \textit{replica} of the mobile YouTube app (since these authors had not been able to modify apps) and explored how the design of the app can both support and undermine users' agency in device use.
HabitLab is an open-source project led by Geza Kovacs that analyses and helps mitigate digital distractions on desktop and mobile~\cite{kovacs_thesis}.
While HabitLab includes a wide range of interventions for \textit{desktop}, the supply of interventions on mobile is limited and is restricted to leveraging APIs provided by the Android operating system~\cite{greaseterminator}.

These examples underline the relative absence of research studies into distraction by mobile apps, due to the difficulty in deploying interventions against distractions on mobile.
The following shows how our implementation of mobile app extensions can help study and tackle distractions in mobile apps.

\textbf{Code-based extensions in the Facebook app.}
Using our framework, we have been able to replicate the arguably most well-known anti-distraction intervention, namely removing the Facebook feed.
We also developed a control condition, by colouring the background of the Facebook feed in light grey, as proposed by Lyngs et al.~\cite{lyngs_i_2020}.

To turn these interventions into app extensions for our Android implementation, we first decoded the Facebook app with \texttt{apktool} into Android bytecode.
Within this code, we searched for references to the News feed and identified which part of the code creates the UI element for the news feed.
We can then insert code into the Facebook app that changes the background colour of the feed, or removes this functionality altogether, see Figure~\ref{fig:facebook_newsfeed}.
There might be simpler ways to develop extensions for the Facebook app, but the current version of \texttt{apktool} fails to decode the Android resource files that describe the UI layout.
However, this current limitation will likely be fixed in a future version of \texttt{apktool}.

\textbf{Resource-based extensions in the Twitter app.}
While the decoding of app layout files failed for the Facebook app, there were no issues with the Twitter app.
Consequently, we were able to develop a range of modification to the Twitter UI, including the removal of the stories bar in the Twitter app, as well as hiding the notification count on Twitter, see Figure~\ref{fig:ptch}.
The notification count acts as a cue for checking notification on Twitter and can initiate a habit loop~\cite{lyngs_self-control_2019}.

\subsection{Privacy}

Previous work has found that end-users wish to have more choice over their privacy in apps~\cite{shklovski_leakiness_2014,van_kleek_better_2017}, but also that most apps do not provide privacy choices. Recent work by Kollnig et al. found that more than 70\% of apps use third-party tracking, but less than 10\% ask for prior user consent, which is required under EU data protection law~\cite{kollnig2021_consent}.
We now describe two categories of extensions that can potentially help researchers develop new user-centric, privacy-enhancing technologies.

\textbf{Easy disabling of certificate pinning.}
A common challenge faced in previous privacy research are the increasing restrictions against the analysis of network traffic~\cite{kollnig2021_consent}.
Recent Android versions do not trust self-signed root certificates, which are key to analysing HTTPS network traffic.
Many popular apps additionally use certificate pinning, which restricts the certificates that an app trusts.
A solution against both these strategies is the use of the tool \texttt{apk-mitm}, which runs on a desktop computer.
Using our proposed implementation of mobile app extensions, we can embed this tool within an extension script, thereby allowing users to run \texttt{apk-mitm} from a mobile device.
While researchers studying certain apps might be able to run \texttt{apk-mitm} on their own computers, this strategy might fail when new privacy-enhancing technologies are deployed to actual end-users.
In this case, our system could help end-users temporarily disable the network security of certain apps to enable new forms of privacy research, e.g. studying how often personal data leaked is unexpectedly to third-party companies by users' favourite apps during a normal day.
Any studies that try to assess individuals' privacy footprint outside the lab would profit from our technology.

\textbf{Unwanted in-app tracking.}
We have developed another range of extensions that tackles the widespread use of tracking capabilities within apps (see Figure~\ref{fig:steps}).
With our architecture, there are at least four different strategies to address tracking in apps: 1) change the tracking SDK configuration within the app manifest, 2) intercept SDK initialisation calls, 3) call SDK configuration functions, and 4) remove tracking host names within apps.
All four strategies would reduce, if not remove, tracking by apps.
The reduction of tracking in apps could enable new forms of privacy, e.g. studying if apps can continue to function without tracking capabilities, and what strategies apps employ when they are cut off from accessing certain kinds of personal information.

\textbf{GPC and DNT: Legally binding opt-out signals from data processing.}
On the web, there is increasing interest in sending legally binding privacy signals to data-collecting companies.
These efforts are spearheaded by the Global Privacy Control (GPC) that encodes `do not sell' (under California's CCPA) or `object to processing' (under the EU's GDPR) signals.
So far, no similar approach exists for mobile apps, despite those being at the forefront of data collection.
App extensions could help address this gap.

\subsection{Child Safety}
\label{sec:kids}

Children are some of the most vulnerable users of digital devices. They also spend ever more time with such devices. This is why it is important to implement thorough protections for children against digital harms. The following two case studies explore how our current implementation of mobile app extensions might help reduce harms for children in apps. These case studies also nicely illustrate the process of developing extensions with our current implementation, as well as current limitations of our approach.

\textbf{Use of location data.}
One common concern among parents is apps' use of location data~\cite{zhao_silly_name_2019,playstore_kids,information_commissioners_office_kids_2021}.
While it is possible to disable the sharing of location data from the device settings, it is possible on many devices to re-enable access to location data just as easily.
Indeed, some apps, such as Pokémon GO, rely on access to location data for their core functionality.
Because of this, we have developed an extension to remove access to location data directly from the app, and instead set the user location to a fixed location within the app code.
If granted the location permission at run-time, this makes apps believe that they have access to the user location, when in fact, user location has been pre-determined and is faked.
The use of this extension might help explore parents' concerns around the use of location data further in future research, by encouraging debate around the development of apps without location access at all.

\textbf{Risk of overspending.}
Another common fears of parents is the \textit{risk of overspending}~\cite{zhao_silly_name_2019}: If children play games on their parents' devices, they may accidentally spend sizeable sums of money with their parents' credit card.
To reduce the potential risks, we have developed a simple extension using our technology that disables the in-app payment functionality of apps.
The extension relies on the fact that apps wanting to invoke in-app payments need to reach out to a System API, which is identified by a unique string, {\small \path{com.android.vending.billing.InAppBillingService.BIND}}.
Our extension searches for this string within the app, and changes all occurrences to something non-existent (e.g. a long random string), making calls to this System API fail.
Crucially, this does not trick apps into thinking that a payment has been made, but rather blocks apps from connecting with the payment provider of the Google Play store.
The result is that payment functionality within apps stops functioning.

Even though app stores have increasingly been rolling out protections for children,
parents still need to trust the platforms, who often face a conflict of interest in retrieving revenues from their app ecosystems and keeping their users safe~\cite{anirudhchi2021,wang_protection_2021,kollnig2022_iphone_android}. Shortcomings with this approach remain, including a continued lack of privacy protections for children despite the GDPR in the EU~\cite{kollnig_before_2021} and COPPA in the US~\cite{reyes_wont_2018}.
With app extensions, rather than parents having to trust mainly in the efforts and good intentions of platforms in keeping their children safe, they could instead put more trust into a reliable third-parties, such as a consumer and children's organisations, who could provide additional protection methods.
These alternative protection methods could additionally include measures to educate children on spending money online.

\subsection{Summary}

Our case studies demonstrate that our implementation can already be used to replicate known interventions against digital distraction and other harms, as well as create new extensions.
This could be rather helpful to conduct more research on mobile.
The development of extensions, as long as they have a clear injection point within the app code and limited complexity, is rather straightforward and therefore fulfils Usability Requirement 2 for researchers and other developers of app extensions.
The development of each discussed extension took us no more than a few hours. The time of extension development decreased as we refined our technical infrastructure, increased our own patch development skills, and had created other patches to model new ones on.
More complex extensions remain more difficult to realise, and will need more research.
We will discuss the implications from our case studies further in the following Section.

\section{Discussion}
\label{sec:discussion}
Motivated by our case studies, we first discuss the limitations and challenges ahead with our implementation and the concept of mobile app extensions more generally.
Despite these limitations, our case studies also arguably demonstrated the promises of mobile app extensions for end-users and researchers alike.
This is why we also discuss some of the promises of a mobile app extension approach, in spite of the challenges that would lie ahead.

\subsection{Challenges Ahead}

As it stands, our implementation has a range of limitations.
This is to be expected given the power imbalance between unaffiliated researchers/developers and makers of the underlying smartphone ecosystem.
It is beyond this present work to address all of these challenges.
Rather, we would like to highlight the principal feasibility of a mobile extension approach, a broad relevance for addressing mobile harms, and the absence of extension methods on mobile devices (as opposed to desktop browsers).

\textbf{Challenges in extension development.}
Our case studies in this section highlight three different approaches to develop extensions: 1) changing the app manifest (e.g. to reduce tracking), 2) changing app resource files (e.g. to remove the stories bar from the Twitter app), and 3) changing the app code (e.g. to remove the Facebook feed, or temporarily trust self-signed certificates).
These three different methods have a varying difficulty in development.
The modification of app manifest files is relatively easy, since the modification options within manifest are usually documented in online documentation.
The modification of app resource files is slightly more complicated because it requires the identification of the right parts of the right layout file, as well as an understanding of Android development.
The changing of app code files is the most complicated, and needs sound programming skills.
However,
if app modification shall be used for research studies, this should not be a massive problem, since researchers could reach out to other researchers who have sufficient programming expertise.

\textbf{Reliance on re-signing.}
At the moment, there exists no native app extension layer integrated into either Android or iOS.
As a result, we make modifications directly to apps to change the default behaviour.
Any changed app with our system must be signed with our own certificate.
This creates the need to install the original version of the app and breaks compatibility with automated updates through the Google Play store.
This can put study participants at a certain risk, if an app is discovered to contain security flaws during extension deployment.
While we believe that this risk is limited since apps run in a sandbox, it would be fairly simple to implement reminders of the need to update within our mobile app and walk participants through the update process.
Indeed, by installing the existing Aurora app store, Android users can already receive update reminders for extended Android apps.

\textbf{Limits of extension tools.}
There are also technical challenges related to decoding apps with \texttt{apktool}.
There are some apps, including the latest version of the Facebook app, that have compatibility issues with \texttt{apktool}, and not all types of modifications can be applied to those apps, see Section~\ref{sec:casestudies}.
The reason for this problem likely lies in the fact that Facebook (and other apps) use obfuscation and other techniques to make the decoding of their apps harder.
While future updates to \texttt{apktool} might overcome some incompatibilities, it will remain a cat-and-mouse game against app developers.
It might be necessary to explore alternative app extension tools in further work.

\textbf{Terms of Services and Research Studies.}
The alteration of the default behaviour of apps might be in conflict with provisions in terms and services of apps or app stores.
At the same time, there exist explicit provisions under international copyright law that allow the alteration of computer code for research purposes (i.e. interoperability clauses that allow to study the functioning of computer programs).
There arguably also exists a strong societal interest in facilitating the freedom of research in studying the harms emerging from algorithmic systems.
As a result of these consideration and existing legal requirements, there is strong reason to believe that the alteration of the default behaviour of programs for the purposes of research studies cannot be outright prohibited by contractual obligations and is legal in many jurisdictions.
A full account of the legal situation is beyond this current paper, in particular because the situation differs between jurisdictions, and needs further work.

\textbf{Potential service bans.}
WhatsApp has been reported to ban users making modifications to their app, by using modification detection mechanisms in their Android app.
Therefore, in a study setting, it is important to do research about any reported bans, and provide users with dummy credentials if necessary to avoid service bans.
It is important to note that service bans, if modifications are made for legitimate research purposes, might actually be against the law, as discussed in the previous paragraph.
We have not found reports about other apps having banned using for the modification of their apps.

\subsection{Better regulation of apps and harms within them}

The evolution of the digital environment is collectively governed by actions taken by agents. Since the start of the World Wide Web (WWW), numerous shifts have taken place to adapt the Internet, and it is still evolving today. 
The evolution of the Internet and the user browsing experience is not centrally-owned by one party. Rather, it relies on the contributions of different stakeholders to refine and improve the experience.
Yochai Benkler famously framed this environment~--~with its unique opportunities and incentives~--~the \textit{networked world}~\cite{benkler_wealth_2006}.
The evolution of the WWW has not halted. There continue to be rich and fundamental debates on how data should be linked in the semantic web \cite{1637364} or how data should be owned and managed with user pods \cite{10.1145/2872518.2890529} inform us that the evolution of how users should interact with the WWW has not halted.

Beyond such fundamental questions around the organisation of the Web, the end-user experience also continues to be refined and face new challenges and harms.
Website developers may curate their websites to their own design, browser developers can design their own browser software, but as long as one browser exists and supports a sufficient network of extension developers, the digital experience of an end-user is not entirely subjugated to the whim of the website and browser developers. 
End-users can unilaterally sample from a broad selection of extensions to customise their browsing experience with immediate effect, with or without prerequisite consent from the developers.
If the developers wish to retaliate (e.g. in the case of ad blockers), they can make corresponding changes to hinder or disable the extensions.
In turn, extension developers usually adapt their extensions, end-users update their extensions, and the whole process repeats. In giving other participants of the digital world the right to modify the default behaviour of software, they quicken the feedback loop around the optimal design of webpages and online platforms. 

The evolution of the digital experience on browsers can be conceptualised as a fast-paced turn-based game to obtain an optimal equilibrium of interaction, whereas that of mobile apps can be conceptualised as a zero-sum game of unilateral benefit and harm for developers and users respectively.
The evolution in mobile apps takes the form of developer-initiated code changes such as through A/B testing to evaluate improvements to their internal metrics, such as increasing engagement and reducing attrition.
In browsers, developers can get quicker feedback on their software and even inspect alternative designs implemented by the crowd. 
In mobile apps, the loop does not progress, and since widespread extensions cannot be adopted by the mass user base, developers are not incentivised to make code changes that harm their self-interests.
Theoretically, users could switch from one app to the other, but in practice, there is often a race to the bottom in terms of apps' practices~\cite{anirudhchi2021,mhaidli_we_2019,hadar_privacy_2018,chitkara_does_2017,balebako_privacy_2014}.

Unlike browsers where actions can be undertaken by different agents, the actions are mostly taken by developers and platforms on mobile devices.
Regulation and mediation channels for users are limited.
In the future, mobile app extensions might help address this gap.

\section{Conclusions}
\label{sec:conclusions}
In this work, we investigated the concept of mobile app extensions, as a means to study and address digital harms within mobile ecosystems.
Through a thorough study of previous literature, we derived key requirements of a mobile app ecosystem, and derived a versatile and ready-to-use implementation for Android.
In a range of case studies, we demonstrated that our current implementation can already help users reduce a range of mobile harms, by adopting known interventions from the web browsers to mobile. 
Within these case studies, we additionally underlined how our method does not only enable end-users to reduce such harms, but also other researchers in pursuing research on harms within apps.
Previously, research efforts struggled to deploy and study interventions in mobile apps, and thus focused on desktop.
Our approach does not aim to eradicate all of these harms, but rather tries to adopt the proven methodology of browser extensions to mobile.
No ready-to-use system had previously existed; we will share our code publicly.

Our current implementation is far from perfect, and has a range of limitations, as we discuss this in detail in Section~\ref{sec:discussion}.
This is to be expected given the steep power imbalance between unaffiliated researchers/developers and makers of the underlying smartphone ecosystem.
Especially in recent years, both Apple and Google have been tightening their control over iOS and Android, respectively, by making it more difficult for researchers to analyse the practices of apps~\cite{kollnig2021_consent,kollnig2022_iphone_android}.
There remains a significant challenge in ensuring ease of use and also a broad range of possible app extensions.

While our \textit{implementation} has limitations, the \textit{concept} of mobile app extensions explored in this work might still be a valuable concept for end-users (to tackle mobile harms) and researchers (to study mobile harms).
Ultimately, if Apple and Google were to integrate extension functionality directly into their mobile operating systems, this would overcome many of the key challenges and help reduce the mobile harms for individuals.
These challenges are similar to web browsers, where a flourishing ecosystem of browser extensions would not be possible without the support of the browser manufacturers, and currently being challenged by the introduction of Manifest V3 by Google in its Chrome browser~\cite{borgolte_understanding_2020}.
At the same time, many previous research studies into digital harms would not have been possible without browser extensions, and so there is arguably a need to have a similar methodology on mobile.

\textbf{Future work.}
In future work, we will study the human aspects of our system in more detail, by conducting surveys and interviews with end-users. Furthermore, we seek to address further any security concerns with our system (which is out of scope for this work, because we envisage that the existing review mechanisms of app stores be used).
There is also further scope for legal analysis around browser and mobile extensions.
However, this is less of interest if modifications are explicitly supported and foreseen by device manufacturers, similar as they are by the developers of desktop browsers.
We currently use a Docker-based implementation of the Mobile App Extension system; in the future, the implementation might be simplified further, by ideally running the whole extension system on users' smartphones~--~this will likely depend on the support by device manufacturers, notably Apple and Google.
Another interesting avenue for future work is the development of improved harms protection mechanisms for children in mobile apps; while Google and Apple have been improving their protections for children, limitations remain; app extensions could allow trustworthy third-parties, such as consumer and children's organisations, to develop protection mechanisms against digital harms for children.


\bibliographystyle{ACM-Reference-Format}
\bibliography{bib}

\end{document}